\documentclass[pra,twocolumn,showpacs,superscriptaddress]{revtex4}

\usepackage{amsfonts,amsmath,amssymb,psfrag}
\usepackage{bm}
\usepackage{txfonts}
\usepackage{color}
\usepackage{subfigure}
\usepackage[dvips]{graphicx}

\newcommand{\DP}[2]{\frac{\bar{\delta}#1}{\bar{\delta}#2}}

\newcommand{\bra}[1]{\langle#1|}

\newcommand{\EQ}[1]{\begin{eqnarray}#1\end{eqnarray}}
\newcommand{\mbf}[1]{\mathbf{#1}}
\newcommand{\av}[1]{\langle#1\rangle}
\newcommand{\myint}[1]{\int d^3 #1\;}

\newcommand{\QQ}{{\cal Q}}
\newcommand{\PP}{ {\cal P}}

\newcommand{\ital}[1] {{\em #1}}
\newcommand{\etal}{{\em et al.}}

\definecolor{myGray}{rgb}{0.6,0.6,0.6}
\definecolor{myGreen}{rgb}{0.1,0.8,0.1}
\definecolor{myBlue}{rgb}{.1,.1,.8}

\newcommand{\old}[1]{}

\begin{document}
\title{Bose-Einstein Condensation from a Rotating Thermal Cloud: \\Vortex Nucleation and Lattice Formation}
\author{A. S. Bradley} 
\affiliation{ARC Centre of Excellence for Quantum-Atom Optics, School of Physical Sciences, University of Queensland, Brisbane, QLD 4072, Australia.}
\author{C. W. Gardiner}
\affiliation{Jack Dodd Centre for Quantum Technology, Department of Physics, University of Otago, PO Box 56, Dunedin, New Zealand.}
\author{M. J. Davis}
\affiliation{ARC Centre of Excellence for Quantum-Atom Optics, School of Physical Sciences, University of Queensland, Brisbane, QLD 4072, Australia.}

\date{\today}

\begin{abstract} 
We develop a stochastic Gross-Pitaveskii theory suitable for the study of Bose-Einstein condensation in a {\em rotating} dilute Bose gas. The theory is used to model the dynamical and equilibrium properties of a rapidly rotating Bose gas quenched through the critical point for condensation, as in the experiment of Haljan {\em et al.} [Phys. Rev. Lett., {\bf 87}, 21043 (2001)]. In contrast to stirring a vortex-free condensate, where topological constraints require that vortices enter from the edge of the condensate, we find that phase defects in the initial non-condensed cloud are trapped {\em en masse} in the emerging condensate. Bose-stimulated condensate growth proceeds into a disordered vortex configuration. At sufficiently low temperature the vortices then order into a regular Abrikosov lattice in thermal equilibrium with the rotating cloud. We calculate the effect of thermal fluctuations on vortex ordering in the final gas at different temperatures, and find that the BEC transition is accompanied by lattice melting associated with diminishing long range correlations between vortices across the system.
\end{abstract}

\maketitle

\section{Introduction}
Rotating Bose-Einstein condensation was first realized by Haljan \etal at JILA~\cite{Haljan2001}, who grew large vortex lattices by evaporatively cooling a rotating thermal cloud. The evaporation is performed in a prolate trap, along the axis of rotation so that escaping atoms make large axial excursions but do not stray far from the rotation axis. Escaping atoms carry away significant energy but little angular momentum, increasing the angular momentum per trapped atom. Using this cooling mechanism, which requires typically $\sim 50$s of evaporative cooling and an extremely cylindrical trap, very large vortex lattices containing more than 300 vortices have been created \cite{Engels2004}. This nucleation mechanism is distinct from more studied vortex formation mechanisms~\cite{Matthews1999,Leonhardt2002,Madison2000,Hodby2002,Abo-Shaeer2001,Inouye2001,Anderson2001a,Dutton2001a,Scherer2007} because, rather than manipulating a condensate to generate vortices, vortices can be formed at the onset of condensation as the rotating gas reaches the critical temperature ($T_C$) for Bose-Einstein condensation. This mechanism thus has connections with the Kibble-Zurek~\cite{Kibble1976,Zurek1985} and Berezinskii-Kosterlitz-Thouless~\cite{Berezinskii1971,Kosterlitz1973,Hadzibabic2006a} mechanisms, as it is fundamentally caused by spontaneous symmetry breaking and thermal fluctuations near a phase transition.

In this work we study the dynamics of vortex lattice formation during rotating evaporative cooling using a stochastic Gross-Pitaevskii equation (SGPE)~\cite{Stoof1999,SGPEI,SGPEII}. Our central aim is to include rotation, thermal fluctuations, and two-body collisions, by separating the system into a condensate band coupled to a (possibly rotating) thermal reservoir. The condensate band is described by a generalized mean field theory known as the truncated Wigner method \cite{Graham1973,Steel1998,Sinatra2001} that includes the projected classical field method~\cite{Davis2001b,Davis2006a,Simula2006a} as a special case in the regime of high temperatures, while the non-condensate band formalism is based on quantum kinetic theory \cite{QKV}. 

We also provide expressions for the dissipation rates of the theory under the condition that the high-energy component is approximately Bose-Einstein distributed, and details of our numerical algorithm that is central to the utility of the SGPE approach. These are the first SGPE simulations of vortex formation in a trapped Bose gas.
\section{System and equations of motion}\label{sec.system}
The trapped system is divided into a condensate band of states with energy beneath the cut-off $E_R$, and the remaining non-condensate band of high-energy states. The formal separation of the Hamiltonian into bands according to energy allows different methods to applied in the two bands. The condensate band is treated using the truncated Wigner phase space method~\cite{Graham1973,Steel1998} which is valid when the modes are significantly occupied~\cite{Blakie2007a}. Within this approach the condensate band includes the condensate and the most highly occupied excitations. The non-condensate band is comprised of many weakly occupied modes and is treated as approximately thermalised; the effect of this band is to generate dissipative terms in the equation of motion for the condensate band. The high-energy states play the role of a grand canonical thermal reservoir for the low energy condensate band. In this section we provide a minimal overview of the formalism and simply state the final equations of motion to be used in this work. Refs.~\cite{SGPEI,SGPEII} contain the essential formalism, and Ref.~\cite{SGPEIII} gives necessary modifications for the rotation system, along with a number of corrections to the derivation of the equations of motion.

\begin{figure}[tbp]
\centering
\includegraphics[width=8cm]{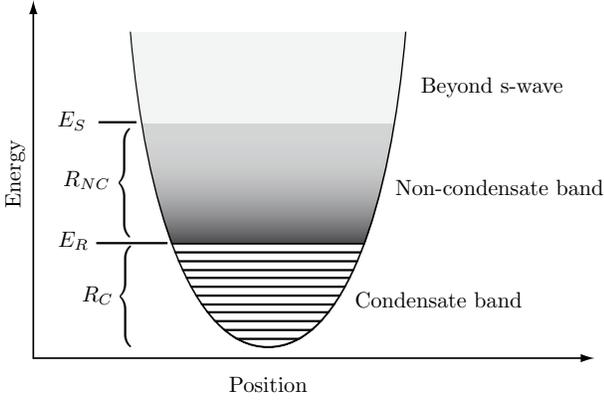}
\caption{Separation of the system into condensate and non-condensate bands. States beneath the energy $E_R$, defined in the rotating frame, form the \ital{condensate band}, while states above $E_R$ form the \ital{non-condensate band} consisting of high-energy, approximately thermalised atoms. In the s-wave scattering regime high-energy states ($E_S\leq \epsilon_{nlm}$) are also eliminated from the theory~\cite{SGPEII}.}
\label{band_schematic}
\end{figure}
\subsection{Hamiltonian and separation of the rotating system}
The cold-collision regime for bosonic atoms of mass $m$ is characterized by the interaction parameter $u=4\pi\hbar^2a/m$, where $a$ is the s-wave scattering length~\cite{Dalfovo1999}. The second-quantized many-body Hamiltonian for a dilute Bose gas confined by a trapping potential $V(\mbf{x},t)$ in the rotating frame defined by angular frequency $\mbf{\Omega}\equiv\Omega\hat{\mbf{z}}$ is $\hat{H}=\hat{H}_{\rm sp}+\hat{H}_I$, where
\begin{equation}\label{Hsp}
\hat{H}_{\rm sp}=\myint{\mbf{x}}\hat{\psi}^\dag(\mbf{x})\left(-\frac{\hbar^2\nabla^2}{2m}-\Omega L_z+V(\mbf{x},t)\right)\hat{\psi}(\mbf{x}),
\end{equation}
and
\begin{equation}\label{ccint}
\hat{H}_I=\frac{u}{2}\myint{\mbf{x}}\hat{\psi}^\dag(\mbf{x})\hat{\psi}^\dag(\mbf{x})\hat{\psi}(\mbf{x})\hat{\psi}(\mbf{x}).
\end{equation}
The orthogonal projectors that separate the system are defined with respect to the single particle basis of the trap. This provides a convenient basis because (a) The many body spectrum becomes well approximated at high-energy by a single particle spectrum, allowing the cutoff to be imposed in this basis \cite{Davis2001b}, and (b) the numerical method used to solve the final equations of motion uses the single particle basis~\cite{Blakie2005a}.

The condensate band projector takes the form
\begin{equation}\label{Pdef1}
\PP\equiv\sum_\mathbf{n}^-|\mathbf{n}\rangle\langle \mathbf{n}|,
\end{equation}
where the bar denotes the high-energy cut-off and $\mathbf{n}$ represents all quantum numbers required to index the eigenstates of the single particle Hamiltonian. To minimize notation we will use $\PP$ both for the operator and for its action on wave functions; the meaning will always be clear from the context. The field operator is decomposed as
\begin{equation}\label{divide_field}
\hat{\psi}(\mathbf{x})=\hat{\phi}(\mathbf{x})+\hat{\psi}_{NC}(\mathbf{x})\equiv\PP\hat{\psi}(\mathbf{x})+\QQ\hat{\psi}(\mathbf{x}),
\end{equation}
where the non-condensate
field $\hat{\psi}_{NC}(\mathbf{x})\equiv\QQ\hat{\psi}(\mbf{x})$ describes the high-energy thermal modes, and the condensate band is described by the field $\hat{\phi}(\mbf{x})\equiv\PP\hat{\psi}(\mbf{x})$. In the spatial representation the eigensolutions of $\hat{H}_{\rm sp}Y_{\bf n}=\epsilon_{\bf n}Y_{\bf n}$, denoted $Y_\mathbf{n}(\mbf{x})\equiv\bra{\mbf{x}}\mathbf{n}\rangle$, generate the action of the projector on an arbitrary wave function $\Psi(\mbf{x})$
\begin{equation}\label{Pdef2}
\PP\Psi(\mbf{x})=\sum_\mathbf{n}^-\;Y_\mathbf{n}(\mbf{x})\myint{\mbf{z}}Y_\mathbf{n}^*(\mbf{z})\Psi(\mbf{z}).
\end{equation}

The condensate band field theory generated by such a decomposition has a nonlocal commutator
\begin{equation}
[\hat{\phi}(\mbf{x}),\hat{\phi}^\dag(\mbf{y})]=\delta_C(\mbf{x},\mbf{y})
\end{equation}
where $\delta_C(\mbf{x},\mbf{y})=\langle \mbf{x}|\PP|\mbf{y}\rangle$. In the harmonic oscillator representation of a one dimensional system, the approximate delta function has a position dependent width, being narrowest in the center of the trap. This reflects the fact that trap wave-functions become smoother near their semi-classical turning points. Note that since $\PP\PP=\PP$, $\PP\hat{\phi}({\bf x})=\hat{\phi}({\bf x})$; put another way, this means that $\delta_C(\mbf{x},\mbf{y})$ is a true Dirac-delta function for any condensate band field. 

\begin{figure}[tbp]
\centering
\includegraphics[width=8.2cm]{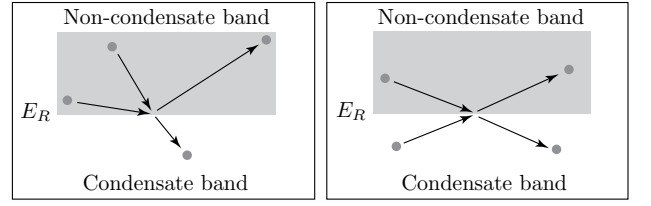}
\caption{Interactions between the condensate and non-condensate bands. In (a) two non-condensate band atoms collide. The collision energy is transferred to one of the atoms with the other passing into the condensate band. In (b) a condensate band atom collides with a non-condensate band atom with no change in condensate band population. The time reversed  processes also occur.}
\label{process_schematic}
\end{figure}

There are two physical processes arising from the interactions between condensate and non-condensate bands, referred to as {\em growth} and {\em scattering} \cite{SGPEI,SGPEII,SGPEIII}, and shown schematically in Figure~\ref{process_schematic}. The growth terms correspond to two-body collisions between condensate and non-condensate atoms whereby particle transfer can take place. The scattering describes inter-band collisions that conserve the populations in each band but involve a transfer of energy and momentum between bands. 
The non-condensate band is described semi-classically in terms of the Wigner function
\begin{equation}
F(\mbf{x},\mbf{K})\equiv\int d^3\mbf{v}\;\av{\hat{\psi}_{NC}^\dag\left(\mbf{x}+\frac{\mbf{v}}{2}\right)\hat{\psi}_{NC}\left(\mbf{x}-\frac{\mbf{v}}{2}\right)}\;e^{i\mbf{K}\cdot\mbf{v}}.
\end{equation}
In this work the trapping potential assumes the cylindrical form
\begin{equation}
V(\mbf{x},t)\equiv\frac{m}{2}(\omega_r^2 r^2+\omega_z^2 z^2),
\end{equation}
where $\omega_r$ and $\omega_z$ are the radial and axial trapping frequencies respectively.
The non-condensate band Wigner function for a Bose gas confined in a cylindrically symmetric trap in equilibrium in the rotating frame at frequency $\Omega$, with chemical potential $\mu$, is
\begin{equation}\label{Fukdef}
F(\mbf{x},\mbf{K})=\frac{1}{\exp{[(\hbar\omega(\mbf{x},\mbf{K})-\mu)/k_BT]}-1},
\end{equation}
where 
\begin{equation}
\begin{split}
\hbar\omega(\mbf{x},\mbf{K})=&\frac{\hbar^2\mbf{K}^2}{2m}-\hbar\mbf{\Omega}\cdot(\mbf{x}\times\mbf{K})+V(\mbf{x}),\\
\label{Krdef}=&\frac{\hbar^2}{2m}(\mbf{K}-m\mbf{\Omega}\times\mbf{x}/\hbar)^2+V_{\rm eff}(\mbf{x}),
\end{split}
\end{equation}
defines the effective potential as $V_{\rm eff}(\mbf{x})=m(\omega_r^2-\Omega^2)^2r^2/2+m\omega_z^2z^2/2$,
and $r=(x^2+y^2)^{1/2}$ is the distance from the symmetry axis of the trap. Since $\mbf{K}_r\equiv\mbf{K}-m\mbf{\Omega}\times\mbf{x}/\hbar$ is the rotating frame momentum, we can work in terms of this variable, defining
\begin{equation}\label{rspec}
\hbar\omega(\mbf{x},\mbf{K})=\hbar\omega_r(\mbf{x},\mbf{K}_r)\equiv \frac{\hbar^2\mbf{K}_r^2}{2m}+V_{\rm eff}(\mbf{x}).
\end{equation}
Hereafter we drop the subscript for brevity since we always describe the system in this frame.
In the semiclassical approximation $F(\mbf{u},\mbf{K})$ is only significant in the non-condensate region $R_{NC}$, which is the region of phase space where $\hbar\omega(\mbf{u},\mbf{K})>E_R$. 
\subsection{Stochastic Gross-Pitaevskii equation} 
Following the theory in Refs.~\cite{SGPEI,SGPEII,SGPEIII}, the equation of motion for the density operator describing the total system
\begin{equation}\label{rhoEOM}
i\hbar\frac{\partial\hat{\rho}}{\partial t}=[\hat{H},\hat{\rho}],
\end{equation}
is reduced to a master equation for the condensate band density operator, which is the trace of $\hat{\rho}$ over the non-condensate band degrees of freedom $\hat{\rho}_C\equiv{\rm tr}_{NC}\left\{\hat{\rho}\right\}$. The master equation is then mapped to a Fokker-Planck equation for the Wigner distribution describing the condensate band. In the truncated Wigner approximation, an equivalent stochastic equation of motion can be obtained that allows the dynamics to be efficiently simulated numerically~\cite{QN}. Formally, there is a correspondence between symmetrically ordered moments of the field operator $\hat{\phi}({\bf x},t)$ and ensemble averages over many trajectories of the projected field $\alpha({\bf x},t)\equiv\bar{\sum}_{\bf n}\alpha_{\bf n}(t)Y_{\bf n}({\bf x})$ that evolves according to the SGPE to be defined below~\cite{SGPEIII}. For example, for the density
\begin{equation}
\langle \alpha^*({\bf x})\alpha({\bf x})\rangle_W=\frac{1}{2}\langle \hat{\phi}^\dag({\bf x})\hat{\phi}({\bf x})+\hat{\phi}({\bf x})\hat{\phi}^\dag({\bf x})\rangle,
\end{equation}
where $\langle\;\;\rangle_W$ denotes stochastic averaging over trajectories of the SGPE. 

To state the SGPE we require some expressions from mean field theory, in particular the GPE obeyed by $\alpha({\bf x})$ in the absence of any thermal atoms in the non-condensate band.
The mean field Hamiltonian for $\alpha({\bf x})$ corresponding to $\hat{H}$ is
\begin{equation}
\begin{split}
H_{\rm GP}=&\int d^3\mbf{x}\;\alpha^*(\mbf{x})\left(-\frac{\hbar^2\nabla^2}{2m}-\Omega L_z+V(\mbf{x})\right)\alpha(\mbf{x})\\\label{GPHdef}
&+\frac{u}{2}\int d^3\mbf{x}\;|\alpha(\mbf{x})|^4,
\end{split}
\end{equation}
with atom number given by $N_{\rm GP}=\int d^3\mbf{x}\;|\alpha(\mbf{x})|^2$. 

The growth terms involve the Gross-Pitaevskii equation found in the usual way as
\begin{equation}
\begin{split}
i\hbar\frac{\partial \alpha({\bf x})}{\partial t}=&\DP{H_{\rm GP}}{\alpha^*(\mbf{x})}\equiv\PP L_{\rm GP}\alpha(\mbf{x})\\
=&\PP\left\{\left(-\frac{\hbar^2\nabla^2}{2m}-\Omega L_z+V(\mbf{x})+u|\alpha(\mbf{x})|^2\right)\alpha(\mbf{x})\right\}
\label{LGPdef}
\end{split}
\end{equation}
that defines the evolution operator $L_{\rm GP}$. To obtain this equation of motion we have used the projected functional calculus~\cite{SGPEII} that plays the same role here as ordinary functional calculus in non-projected field theory. 

The scattering process couples to the divergence of the condensate band current
\begin{equation}
\begin{split}
\mbf{j}_{\rm GP}(\mbf{x})\equiv& \frac{i\hbar}{2m}\Big([\nabla\alpha^*(\mbf{x})]\alpha(\mbf{x})
-\alpha^*(\mbf{x})\nabla\alpha(\mbf{x})\Big)\\&-(\mbf{\Omega}\times\mbf{x})\;|\alpha(\mbf{x})|^2,
\end{split}
\end{equation}
where the second line is a rigid body rotation term arising from the transformation to the rotating frame. The limit ${\bf j}_{\rm GP}({\bf x})=0$ gives the laboratory frame velocity field ${\bf v}=\mbf{\Omega}\times\mbf{x}$, corresponding to the irrotational system mimicking rigid body rotation. Persistent currents around the vortex cores are responsible for maintaining irrotationality.
\subsubsection{Full non-local SGPE}
We finally arrive at
the Stratonovich stochastic differential equation (SDE)~\cite{SGPEIII}
\begin{equation}
 \begin{split}
(S)d\alpha(\mbf{x},t)=&-\frac{i}{\hbar}\PP L_{\rm GP}\alpha(\mbf{x})dt\\
&+\PP\Bigg\{\frac{G(\mbf{x})}{k_BT}(\mu-L_{\rm GP})\alpha(\mbf{x})dt+dW_G(\mbf{x},t)\\
&+\frac{i\hbar\alpha(\mbf{x})}{k_BT}\int d^3\mbf{x}^\prime\;M(\mbf{x}-\mbf{x}^\prime) \nabla\cdot \mbf{j}_{\rm GP}(\mbf{x}^\prime)dt\\
&+i\alpha(\mbf{x})dW_{M}(\mbf{x},t)\Bigg\}\label{SGPEnonloc},
\end{split}
\end{equation}
where the growth noise is complex, and the scattering noise is real. The noises are Gaussian, independent of each other, and have the non-zero correlations 
\EQ{\label{Gnoise}
\langle dW_G^*(\mbf{x},t)dW_G(\mbf{x}^\prime,t)\rangle&=&2G(\mbf{x})\delta_C(\mbf{x},\mbf{x}^\prime)dt,\\
\label{Mnoise}\langle dW_{M}(\mbf{x},t)dW_{M}(\mbf{x}^\prime,t)\rangle&=&2M(\mbf{x}-\mbf{x}^\prime)dt.
}
The growth and scattering amplitudes, $G(\mbf{x})$ and $M(\mbf{x})$, are calculated in Appendix \ref{app:rates} for the case where the non-condensate band is described by a thermal Bose-Einstein distribution. 

The two validity conditions for this equation are (i) {\em high temperature}: $\hbar\omega\ll k_BT$ where $\omega$ is the characteristic system frequency, and (ii) {\em significant occupation} of modes in the condensate band, necessitated by the truncated Wigner method.
The first line of Eq.~\ref{SGPEnonloc} is the PGPE developed by Davis {\em et al}~\cite{Davis2001b} that, in essence, is the GPE formally confined to a low energy subspace of the condensate band. The condensate growth terms of the second line are closely connected to phenomenological dissipative theories of BECs based on the GPE~\cite{Choi1998,Penckwitt2002,Tsubota2002} but here have a specific form given in Appendix \ref{app:rates}. The remaining terms describe scattering processes that transfer energy and momentum between the two bands; these terms where first derived in Ref.~\cite{SGPEII}, and have not appeared in previous SGPE theories. 

Irrespective of the form of $G({\bf x})$ and $M({\bf x})$, the equilibrium state for the Wigner distribution of the system has the grand canonical form~\cite{SGPEII}
\begin{equation}
W_s\propto \exp{\left(\frac{\mu N_{\rm GP}-H_{\rm GP}}{k_BT}\right)},
\end{equation}
that is generated by the stochastic evolution of Eq.~(\ref{SGPEnonloc}). Solutions of the SGPE also known to satisfy a set of generalized Ehrenfest relations which in the steady state reduce to a statement of the fluctuation-dissipation theorem for the grand canonical Bose gas~\cite{Bradley2005b}.
\subsubsection{Simple growth SGPE}
A simpler equation that reaches the same equilibrium state can be obtained by noting that the scattering terms in Eq.~(\ref{SGPEnonloc}) do not directly change the condensate band population. Such terms have been included in a kinetic theory of Bose-Einstein condensation~\cite{QKIV}, where they were found to play a role in the early stages of condensate growth by causing atoms in many modestly occupied states to merge more rapidly into a single highly occupied state which then dominates the growth process. Within the SGPE theory, the scattering terms generate an effective potential from density fluctuations. This can be seen by noting that upon neglecting the growth and scattering terms, that are relatively small compared to the mean field evolution, the continuity equation for the condensate band gives $\nabla\cdot\mbf{j}_{\rm GP}(\mbf{x})=-\partial n_{\rm GP}(\mbf{x})/\partial t$. Consequently, since $M(\mbf{x})$ is purely real, the deterministic part of the scattering terms generate an effective potential that will tend to smooth out density fluctuations.

Upon neglecting these terms the resulting stochastic equation is much easier to integrate numerically as it does not require the non-local noise correlations of the full equations. The simulations in this work are performed by numerically integrating the simple growth SDE
\begin{equation}
\begin{split}
d\alpha(\mbf{x})=&-\frac{i}{\hbar}\PP L_{\rm GP}\alpha(\mbf{x})dt\\
&\PP\left\{\frac{\gamma}{k_BT}(\mu-L_{\rm GP})\alpha(\mbf{x})dt+dW_\gamma(\mbf{x},t)\right\},
\label{SGPEsimp}
\end{split}
\end{equation}
using the mixed quadrature pseudo-spectral method detailed in Appendix \ref{app:numerics}. The rate $\gamma$ is given by Eq.~(\ref{approxG}), and the noise is complex Gaussian with non-vanishing correlator
\begin{equation}
\label{GamCor}
\langle dW_\gamma^*(\mbf{x},t)dW_\gamma(\mbf{x}^\prime,t)\rangle=2\gamma\delta_C(\mbf{x},\mbf{x}^\prime)dt.
\end{equation}
\section{Simulations of Rotating Bose-Einstein condensation}\label{simsec}

\begin{figure}[tbp]
\centering
%
\includegraphics[width=8.5cm]{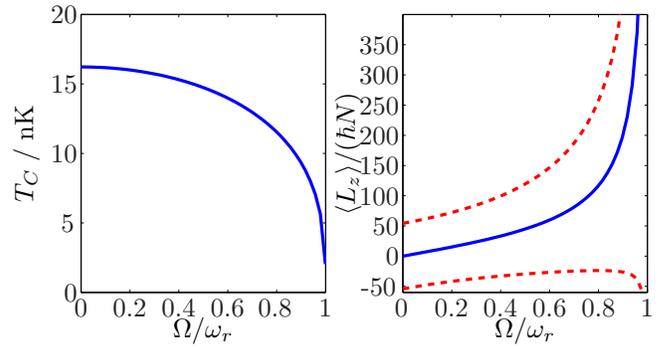}%
\caption{(a) Variation of ideal gas BEC transition temperature with angular frequency of rotation $\Omega$, for $N=1.3\times 10^5$ atoms. (b) Angular momentum per particle (solid line), for the same number of atoms at temperature $T=T_c$, with $\langle L_z\rangle/(\hbar N)\pm \sigma(L_z)/\hbar$ (dashed lines) representing the size of fluctuations. Both the angular momentum per particle and its fluctuations diverge at at the critical rotation frequency.}
\label{fig:thermalWig}
\end{figure}
\begin{figure*}[tbp]
\centering
\includegraphics[width=15.5cm]{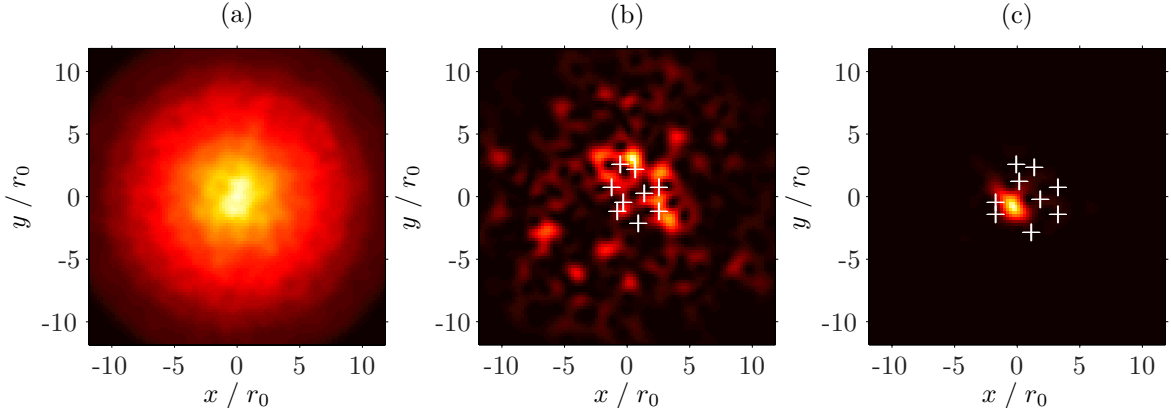}%

\caption{(color online)
The Wigner function for the rotating ideal Bose gas used as the initial state for our simulations of rotating Bose-Einstein condensation. Parameters are $T_0=12{\rm nK}, \Omega_0=0.979\omega_r, \mu_0=0.5\epsilon_{000}$. (a) Mean density $\langle \hat{\psi}^\dag(\mbf{x})\hat{\psi}(\mbf{x})\rangle$ computed by averaging over 5000 samples of the Wigner function. A single sample is shown in (b) with vortices ($+$ have positive angular momentum) shown for $r_v<3 r_0$. (c) Condensate wave-function according to the Penrose-Onsager criterion. The one body density matrix used here is constructed from $500$ samples. Due to the near-degeneracy of angular momentum modes averaging must be taken over many samples to completely remove all vortices from the proto-condensate. In this example the proto-condensate contains $\sim 50$ atoms and many vortices are visible. In all plots the colormap is scaled to represent the entire range of density in the data.
}
\label{fig:thermalWig2}
\end{figure*}
\begin{figure}[tbp]
\centering
\includegraphics[width=7.5cm]{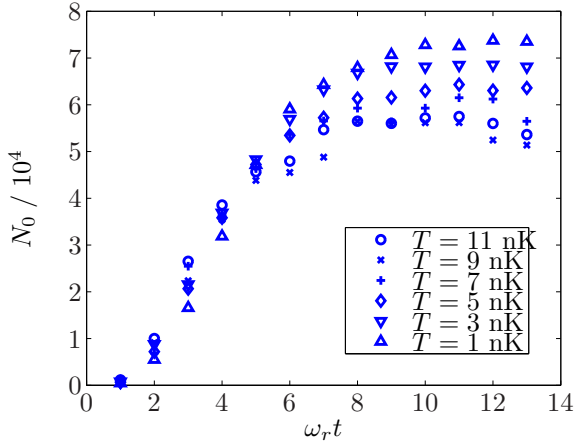}%

\caption{Condensate number $N_0$ during a quench from $[\mu_0,T_0]=[0.5\epsilon_{000},12{\rm nK}]$ to $[\mu,T]=[3.5\epsilon_{000},1{\rm nK}]$ at fixed rotation frequency $\Omega_0=0.979\omega_r$. The condensate number is found by diagonalising the one-body density matrix found from short time averaging single runs of the SGPE (see text). The final condensate number is suppressed at higher temperature.}
\label{fig:N0vst}
\end{figure}
The central question of interest for this work regards the formation of vortices during condensation. In particular, does a vortex-free condensate form and then vortices enter it from the edge, or does condensation proceed into a state with vortices already present?  That this is non-trivial to answer is easily seen from the single particle energy spectrum of the cylindrically symmetric harmonic trap in three dimensions
\begin{equation}
\epsilon_{nlm}=\hbar\omega_r(2n+|l|+1)-\hbar\Omega l+\hbar\omega_z(m+1/2),
\end{equation}
where $n, l, m$ are the radial, angular and axial quantum numbers. In the absence of interactions BEC will form in the ground state mode with energy $\epsilon_{000}=\hbar\omega_r+\hbar\omega_z/2$. Vortices arise from occupation of states with nonzero angular momentum and the positive angular momentum part of the spectrum behaves as $\hbar(\omega_r-\Omega)l$ leading to near-degeneracy of positive angular momentum modes as $\Omega\to\omega_r$. States with angular momentum $\langle \hat{L}_z\rangle=\hbar l>0$ are increasingly easy to excite as the rate of rotation increases and in the interacting gas there is a trade-off between the interaction and rotational contributions to the energy due to the presence of a vortex that is known to favor the creation of vortices in the condensate once the system is rotating faster than the critical angular frequency, $\Omega_c$, for energetic stability of a vortex~\cite{Fetter2001}.
For a rapidly rotating BEC transition ($\Omega\gg\Omega_c$), we can expect that vortices will play a dominant role at all stages in the BEC growth process. It is then possible that {\em atoms condense into a vortex-filled state}. In general the vortex configuration can be disordered. It is not required, for example, that the vortices minimize the system energy by forming a regular Abrikosov lattice at all stages of condensate growth.

\subsection{Properties of the rotating ideal Bose gas and the non-condensate band}
An important effect of rotation in Bose-Einstein condensation is the reduction of $T_C$~\cite{Stringari1999,Coddington2004a}, that, along with the slow one-dimensional evaporative cooling, accounts for the long evaporation time in the JILA experiments. The transition temperature for the ideal Bose gas is modified by the effective radial frequency in the rotating frame, $\omega_\perp\equiv\sqrt{\omega_r^2-\Omega^2}$, so that
\begin{equation}\label{Tc}
T_C=T_C^{o}\left(1-\frac{\Omega^2}{\omega_r^2}\right)^{1/3},
\end{equation}
where $T_C^{o}=0.94\hbar(\omega_r^2\omega_z)^{1/3}N^{1/3}/k_B$ is the nonrotating transition temperature for $N$ trapped atoms. 
\par
In this work we model the non-condensate region assuming the high-energy atoms are quickly thermalised into a Bose-Einstein distribution, that we approximate as non-interacting since the density is usually small at high energies. We can find analytical expressions for the properties of the noninteracting gas, including the effect of the cutoff, by defining an {\em incomplete Bose-Einstein function} 
\begin{equation}
\begin{split}
 g_\nu(z,y)&\equiv\frac{1}{\Gamma(\nu)}\int_y^\infty dx\;x^{\nu-1}\sum_{l=1}^\infty(ze^{-x})^l,\\
&=\sum_{l=1}^\infty\frac{z^l}{l^\nu}\frac{\Gamma(\nu,yl)}{\Gamma(\nu)},
\end{split}
\end{equation}
where $\Gamma(\nu,x)\equiv \int_x^\infty dy\;y^{\nu-1}e^{-y}$ is the incomplete Gamma function. In analogy with the reduction to an ordinary Gamma function $\Gamma(\nu,0)=\Gamma(\nu)$, we have $g_\nu(z,0)=g_\nu(z)\equiv\sum_{l=1}^\infty z^l/l^\nu$, reducing to the ordinary Bose-Einstein function. We then find for the non-condensate band density, for example
\begin{eqnarray}
\begin{split}
n_{NC}({\bf x})=&\int_{R_{NC}} \frac{d^3{\bf k}}{(2\pi)^3}F({\bf x},{\bf k})\\
=&\frac{1}{(2\pi)^3}\sum_{n=1}^\infty e^{\beta\left[\mu-V_{\rm eff}({\bf x})\right]}\int_{K_R({\bf x})}^\infty dk\;k^2\;e^{-n\beta\hbar^2k^2/2m},
\end{split}
\end{eqnarray}
where $\hbar^2 K_{R}({\bf x})^2/2m={\rm max}\left\{E_R-V_{\rm eff}({\bf x}),0\right\}$. Changing integration variable to $y=n\beta\hbar^2k^2/2m$ gives
\begin{eqnarray}
\begin{split}
n_{NC}({\bf x})=g_{3/2}\left(e^{\beta\left[\mu-V_{\rm eff}({\bf x})\right]},\beta\hbar^2 K_{R}({\bf x})^2/2m]\right)/\lambda_{\rm dB}^3,
\end{split}
\end{eqnarray}
where $\lambda_{\rm dB}=\sqrt{2\pi\hbar^2/mk_BT}$ is the thermal de Broglie wave length. Setting $E_R=0$, we recover the standard form for the semi-classical particle density of the ideal gas~\cite{Dalfovo1999}. The total number in the non-condensate band is given by 
\begin{equation}\label{NNC}
N_{NC}=g_3\left(e^{\beta\mu},\beta E_R\right)/(\beta\hbar\bar{\omega})^3,
\end{equation}
where $\bar{\omega}=(\omega_z\omega_\perp^2)^{1/3}$ is the geometric mean frequency in the rotating fame. In this way the usual semi-classical expressions can be generalized to include a cutoff in terms of the incomplete Bose-Einstein function.
\begin{figure*}[tbp]
\centering
\includegraphics[width=15.5cm]{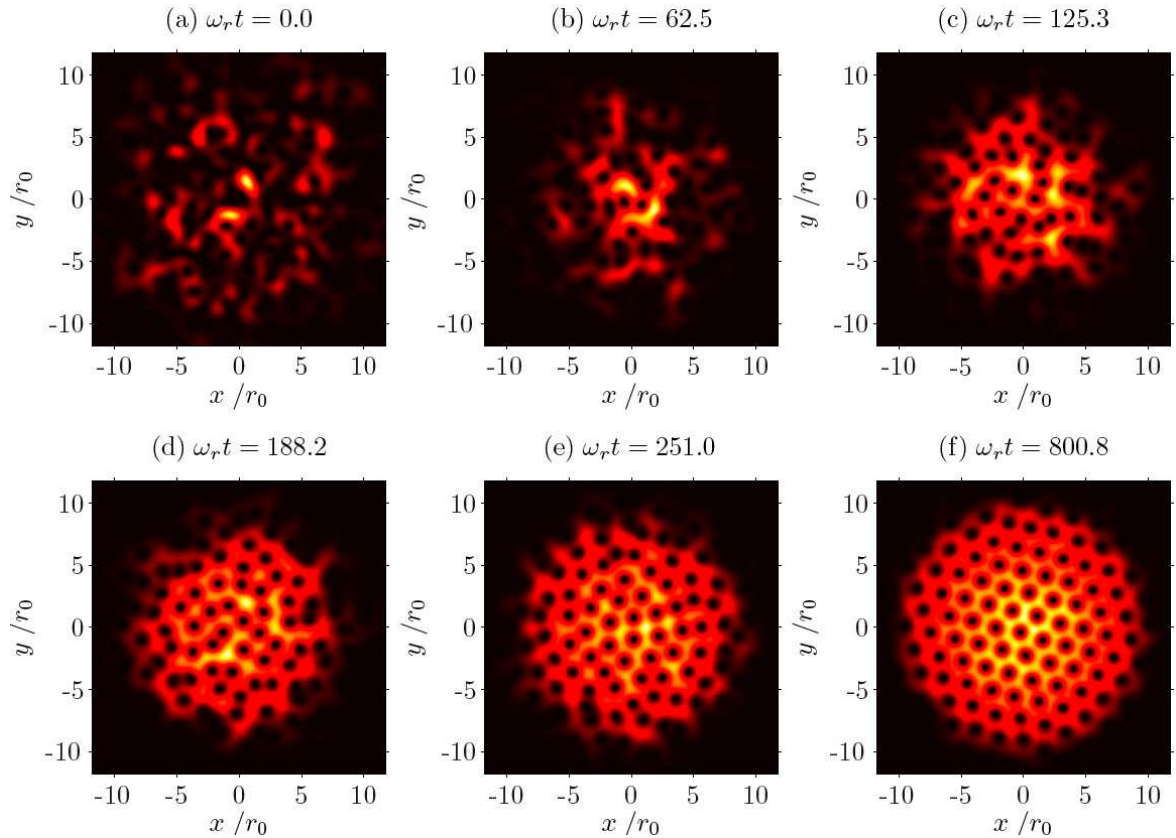}%
\caption{(color online) Condensate band density for a single trajectory of the SGPE. (a) Initial state for $1.3\times 10^6$ $^{87}{\rm Rb}$ atoms at $T_0=12 {\rm nK}$, with $\mu_0=0.5\epsilon_{000}$, and $\Omega_0=0.979\omega_r$. At time $t=0$ the non-condensate band is quenched to $T=1 {\rm nK}$, $\mu=3.5\epsilon_{000}$, and $\Omega=\Omega_0$. (b)-(d) The condensate band undergoes rapid growth. (e)-(f) At this low temperature the vortices assemble into a regular Abrikosov lattice.}
\label{single_trajectory_sgpeRBEC}
\end{figure*}
\begin{figure*}[tbp]
\centering
\includegraphics[width=15.5cm]{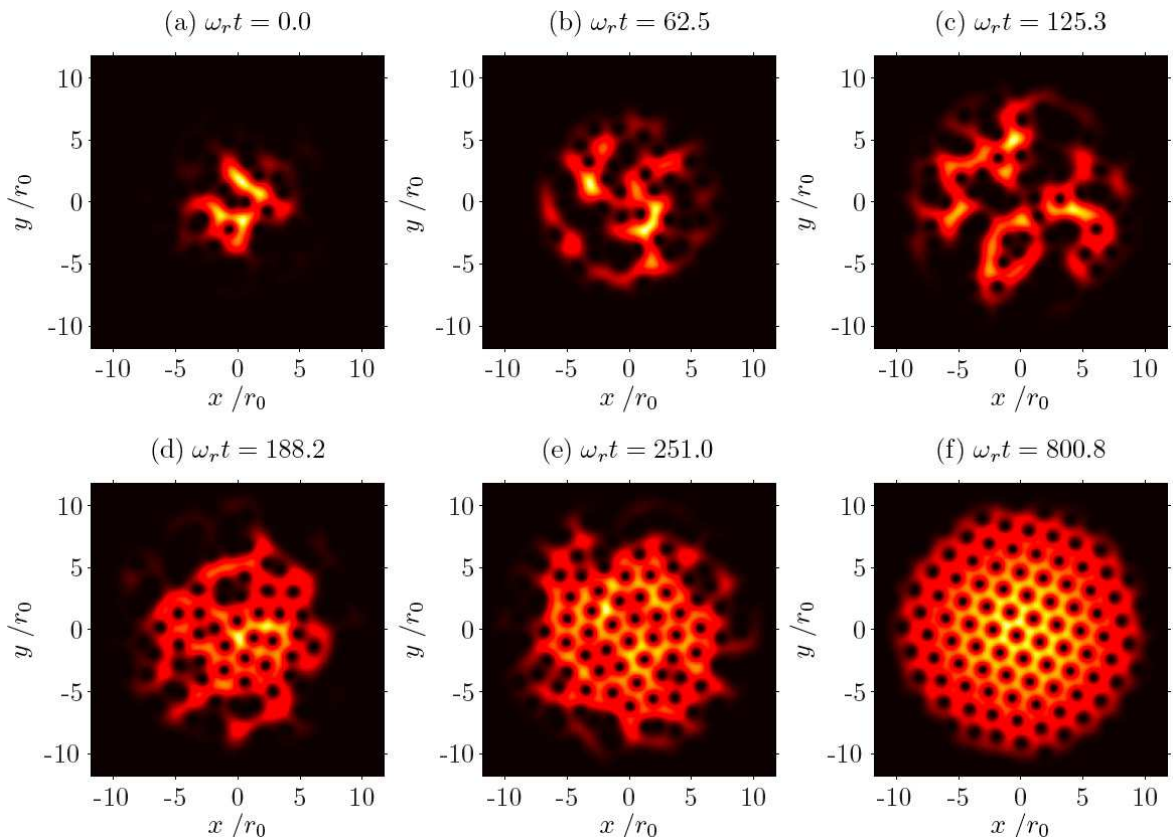}%
\caption{(color online) Condensate mode calculated from the Penrose-Onsager criterion via short time averaging, for the simulation data of Fig~\ref{single_trajectory_sgpeRBEC}. (a) Initial state. (b)-(d) After the quench the condensate undergoes a rapid growth into a highly disordered phase. (e)-(f) The system undergoes a prolonged stage of ordering into a periodic vortex lattice. At this temperature the condensate and condensate band are almost identical in the final equilibrium state.}
\label{condBand_single_trajectory_sgpeRBEC}
\end{figure*}

\begin{figure*}[tbp]
\centering
\includegraphics[width=15.5cm]{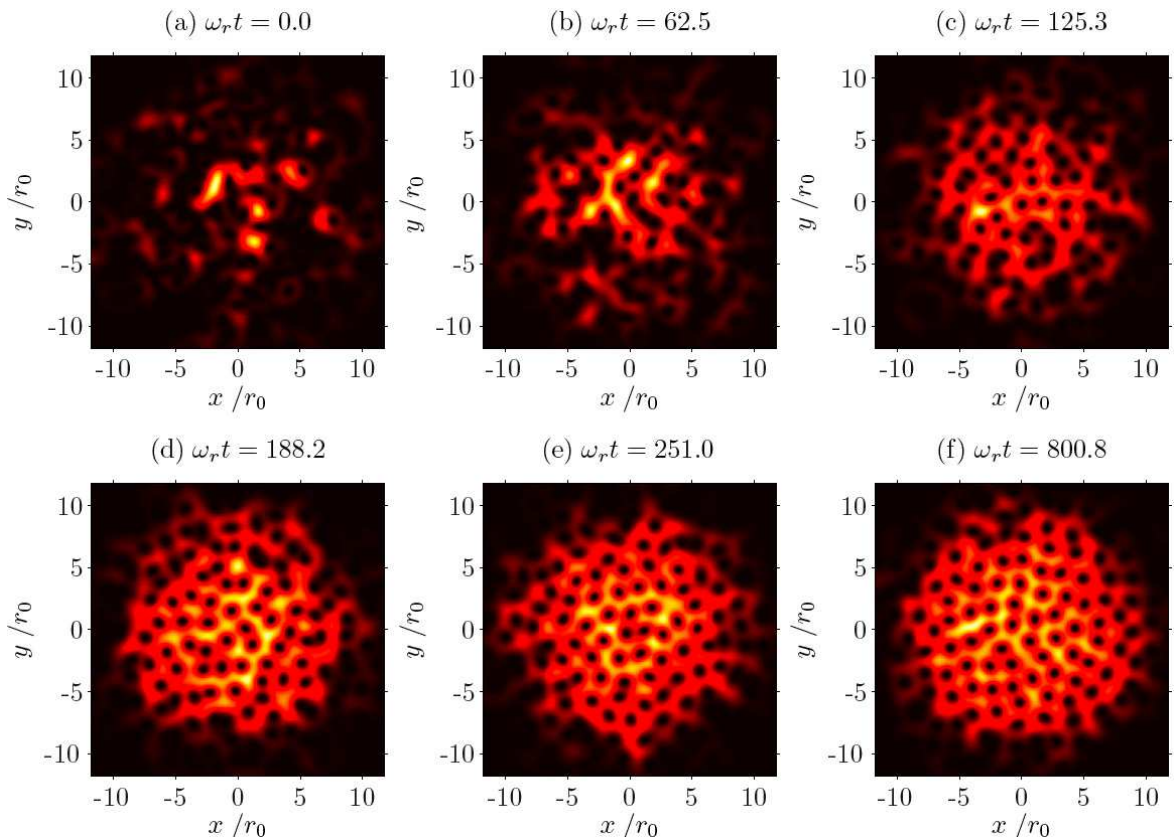}%
\caption{(color online) Condensate band density for a quenched rotating Bose gas. (a) Initial state for $1.3\times 10^6$ $^{87}{\rm Rb}$ atoms at $T_0=12 {\rm nK}$, with $\mu_0=0.5\epsilon_{000}$, and $\Omega=0.979\omega_r$. At time $t=0$ the non-condensate band is quenched to $T=11 {\rm nK}$, $\mu=3.5\epsilon_{000}$, and $\Omega=\Omega_0$. (b)-(d) The condensate band undergoes rapid growth. (e)-(f) After growth, vortex order is suppressed by thermal fluctuations, frustrating the formation of a periodic vortex lattice seen at the much lower temperature of Fig.~\ref{single_trajectory_sgpeRBEC}.}
\label{second_single_trajectory_sgpeRBEC}
\end{figure*}

\begin{figure*}[tbp]
\centering
\includegraphics[width=15.5cm]{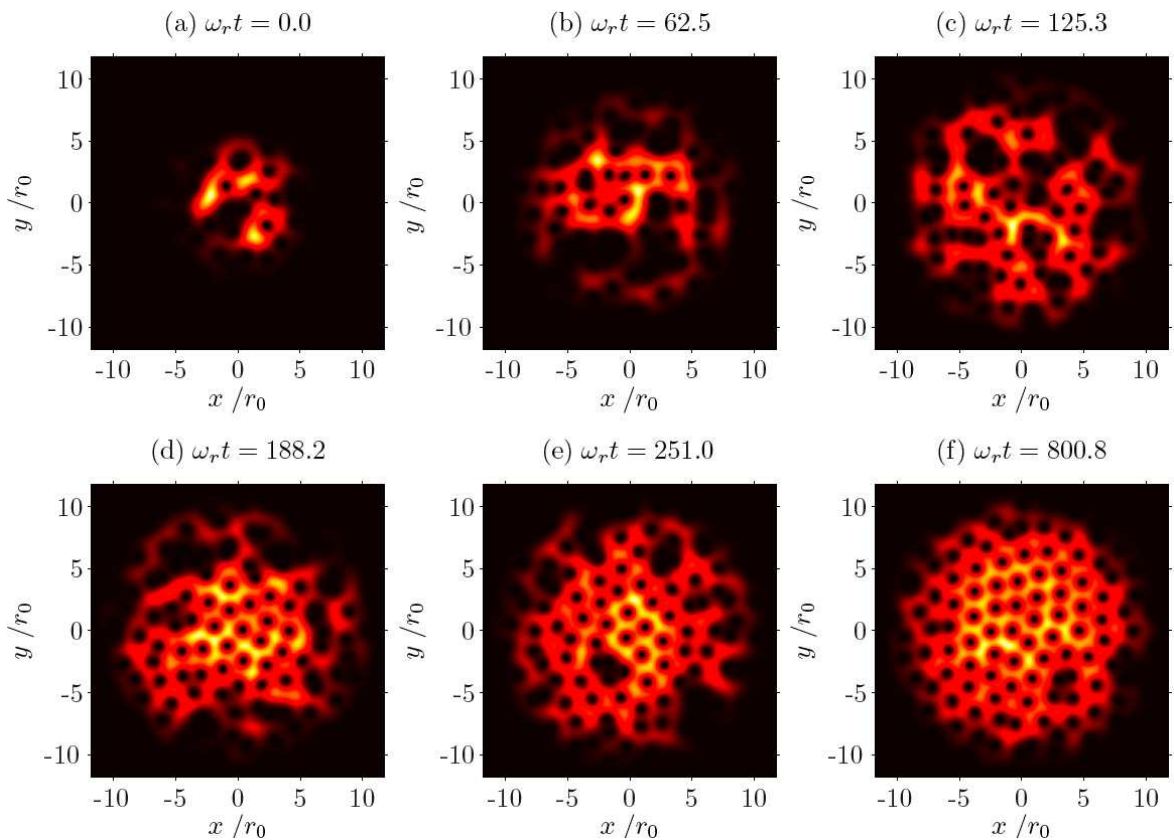}%
\caption{(color online) Condensate mode calculated from the Penrose-Onsager criterion via short time averaging, for the simulation data of Fig~\ref{second_single_trajectory_sgpeRBEC}. (a) Initial state. (b)-(d) After the quench the condensate growth into a highly disordered vortex configuration. (e)-(f) Suppression vortex order is reflected in the disordered condensate wave-function. At this higher temperature the condensate and condensate band are distinct, reflecting the disorder and the increased thermal fraction in the condensate band.}
\label{condBand_second_single_trajectory_sgpeRBEC}
\end{figure*}
The angular momentum and its fluctuations can also be calculated. We now write $\langle A\rangle=(2\pi)^{-3}\int d^3\mbf{x}\int d^3\mbf{k}\; A(\mbf{x},\mbf{k})F(\mbf{x},\mbf{k})$, and the variance per particle as $\sigma(A)^2=\langle A^2\rangle/N_{NC} -(\langle A\rangle/N_{NC})^2$. In rotating frame coordinates the angular momentum is $L_z=\hbar(xk_y-yk_x)+m\Omega(x^2+y^2)$, which in equilibrium gives the rigid body relation between angular momentum and rotational inertia $\langle L_z\rangle =m\Omega\langle x^2+y^2\rangle$. From this expression we find the angular momentum is
\begin{equation}
\langle L_z\rangle=2\hbar\frac{\Omega}{\omega_\perp}\frac{g_4(e^{\beta\mu},\beta E_R)}{(\beta\hbar\omega^\prime)^4},
\end{equation}
where $\omega^\prime=(\omega_z\omega_\perp^3)^{1/4}$.
Similarly, the mean squared angular momentum is
\begin{equation}
\langle L_z^2\rangle=\frac{2\hbar^2g_5(e^{\beta\mu},\beta E_R)}{(\beta\hbar\tilde{\omega})^5}\left(1+\frac{4\Omega^2}{\omega_\perp^2}\right),
\end{equation}
where $\tilde{\omega}=(\omega_z\omega_\perp^4)^{1/5}$. In Figure~\ref{fig:thermalWig} we give an example of the the variation of $T_C$, and the angular momentum per particle together with its fluctuations as a function of the angular frequency of rotation. The suppression of $T_C$ by the rotation is seen to be associated with diverging angular momentum and angular momentum fluctuations near the critical frequency for stable rotation in the harmonic trap. 

\subsection{Numerical procedure}
The long cooling time and the large changes in length scales in the axial and radial dimensions during cooling makes modeling the full condensation dynamics challenging. However, as the gas cools and spins up, the system becomes progressively more two-dimensional. 
We model the system by considering the two dimensional projected subspace of the quasi-two dimensional system. We thus use the SGPE of Eq.~\ref{SGPEsimp}
with rescaled interaction parameter $ u_{2D}=4\pi\hbar^2 a/m L_z$, where in this work we take the thickness through the oblate system as $L_z=\sqrt{2\pi\hbar/m\omega_z}$, obtained by projecting onto the ground state of the $z$-dimension. Note that a consequence of the approximation $G({\bf x})\approx \gamma$ is that the growth rate is unchanged by reduction to the two dimensional effective equation. 

We take as our starting point the gas after it has been evaporatively cooled to a rapidly rotating state above the transition temperature. We then simulate the dynamics of a rapid quench below threshold. 
The procedure is summarized as follows:
\begin{itemize}
\item[(i)] Initial states are generated by sampling the grand canonical ensemble for the ideal Bose gas for initial parameters $T_0$, $\Omega_0$ and $\mu_0$. 
\item[(ii)] To approximate a rapid quench, the temperature and chemical potential of the non-condensate band are altered to the new values $T<T_0$ and $\mu>\mu_0$, while the rotation of the cloud $\Omega$ is held fixed at $\Omega_0$, consistent with axial evaporation. The condensate band field is then evolved according to the SGPE. 
\end{itemize}
Specifically, we aim to model a possible run of the JILA experiment consisting of a system of $^{87}{\rm Rb}$ atoms held in a magnetic trap with frequencies $(\omega_z,\omega_r)=2\pi(5.3,8.3){\rm Hz}$, initially at temperature $T_0 = 12\;{\rm nK}$ with chemical potential $\mu_0=0.5\epsilon_{000}$, and rotating with angular frequency $\Omega_0=0.979\omega_r$. We examine a range of final quench temperatures $T=[1, 2, 3, 4, 5, 7, 9, 11]\;{\rm nK}$ with constant chemical potential $\mu=3.5\epsilon_{000}$. We note the ratio $\Omega/\mu= 0.28\ll 1$, where a value approaching unity indicates the onset of the mean field quantum Hall regime. The ratio $\omega_z/\sqrt{\omega_r^2-\Omega^2}\sim 3.13 \gg 1$, indicates the system is oblate but not highly two-dimensional. We have found that simulations for parameters that are highly oblate are very challenging numerically, so we consider as our first application of the method these more modest parameters. For the results presented here we use a cut-off value of $E_R=5\hbar\omega_r$, giving a condensate band containing $291$ single particle states. To check the cut-off independence of our results we have also carried out simulations for $E_R=6\hbar\omega_r$ (not shown here). We find qualitative agreement for single trajectory dynamics and quantitative agreement (at the level of a few percent) for the ensemble averaged condensate fraction shown in Figure~\ref{fig:condFrac}. We are thus confident that our results are cut-off independent. For computational reasons we choose constant dimensionless damping $\hbar\gamma/k_BT=0.01$ for all simulations. Although we have derived a form for the damping rate $\gamma$ given in Appendix \ref{app:rates}, the values of $\gamma$ arising from Eq.~(\ref{approxG}) for our parameters are orders of magnitude smaller then our chosen value, so we have chosen a fixed damping such that the growth rate is comparable to the JILA experimental time of 50s. This affects the specific time-scale for growth, but not the qualitative features of the simulations which merely require that the damping is much slower than any other time-scale of the system. We emphasize that the equilibrium properties are independent of $\gamma$.

Since our numerical simulation method employs a projected eigenbasis of the associated linear Schr\"{o}dinger problem, the initial Bose-Einstein distribution is easily sampled. 
The grand canonical Wigner distribution for a non-interacting trapped Bose gas distributed over a set of single particle states with frequencies $\omega_{nl}$ is
\begin{equation}\label{eq:wiginit}
W(\{\alpha_{nl},\alpha_{nl}^*\})=\prod_{nl}^- \frac{1}{\pi (N_{nl}+1/2)}\exp{\left\{-\frac{|\alpha_{nl}|^2}{N_{nl}+1/2}\right\}},
\end{equation}
where $\alpha(\mbf{x})=\bar{\sum}_{nl}\alpha_{nl}Y_{nl}(\mbf{x})$, the modes are defined in Eq.~\ref{RFeigs}, and 
\begin{equation}
N_{nl}=\frac{1}{e^{(\hbar\omega_{nl}-\mu_0)/k_BT_0}-1}.
\end{equation}
The energy cut-off is denoted by the upper limit of the product notation. The single particle spectrum of the condensate band is restricted by the cutoff to
\begin{equation}
\hbar\omega_{nl}\equiv\hbar\omega_r(2n+|l|-\Omega l+1)\leq E_R.
\end{equation}
The distribution is sampled as $\alpha_{nl}=\sqrt{N_{nl}+1/2}\;\eta_{nl}$, where $\eta_{nl}$ are complex Gaussian random variates with moments $\langle\eta_{nl}\eta_{mq}\rangle=\langle\eta_{nl}^*\eta_{mq}^*\rangle=0$, and $\langle\eta_{nl}^*\eta_{mq}\rangle=\delta_{nm}\delta_{lq}$. We simulate the time dynamics of the quenched system using the numerical algorithm described in Appendix \ref{app:numerics}. 

We note here that application of the Penrose-Onsager criterion for obtaining the BEC~\cite{Penrose1956,Blakie2005a} is hindered in this system because each run of the SGPE theory leads to vortices in different locations, and ensemble averaging the trajectories generates spurious results for $\langle \hat{\psi}^\dag(\mbf{x})\hat{\psi}(\mbf{x}^\prime)\rangle$. Specifically, the condensate number is significantly reduced below the condensate number that we find using a different technique of short-time averaging single trajectories. Generalizing the Penrose-Onsager criterion for BEC to multi-mode mixed states is not pursued in this work. Here we have taken the view that a given trajectory samples a {\em sub-ensemble} of the full range of quantum evolution, consistent with spontaneously broken symmetry in the initial conditions. We present single trajectory results of our simulations of the SGPE that give a qualitative indication of what could be seen in individual experimental measurements of the atomic density time series. 
\subsection{Initial states and dynamics}
In Figure~\ref{fig:thermalWig2} we show the density of the condensate band for the thermal gas above the BEC transition. Individual samples of the Wigner function are seen to contain phase singularities. We also construct and diagonalize the one body density matrix, according the the Penrose-Onsager criterion for BEC. By averaging $500$ samples from the Wigner function we find that the largest eigenvalue is $N_0\sim 50$ and the associated eigenvector contains several phase singularities. Using $5000$ samples washes out the singularities and gives the noninteracting ground state; the point here is that the high degree of degeneracy means that we must average over very many samples to remove the singularities. Put another way, this means that individual trajectories always contain many vortices, and the motion of these vortices destroys long range coherence associated with BEC.

In Figure~\ref{fig:N0vst} we plot the condensate number as a function of time after the quench for a range of different quench temperatures. The condensate is calculated by constructing the one body density matrix for the sub-ensemble via short time averaging individual trajectories~\cite{Blakie2005a}. We have found that by averaging over $50$ samples taken uniformly over an interval of $2.5$ radial trap periods, we obtain sufficiently good statistics to determine the condensate number $N_0$, given by the largest eigenvalue of $\langle\hat{\psi}^\dag({\bf x})\hat{\psi}({\bf x}^\prime)\rangle$. The main point to note from this result is that at higher temperatures the absolute condensate number is slightly suppressed. However, as we will see in the next section, the condensate {\em fraction} is strongly suppressed, due to the increasing atom number in the non-condensate band for higher temperatures.   

A typical time series is shown in Fig.~\ref{single_trajectory_sgpeRBEC}, for the lowest temperature considered in Fig.~\ref{fig:N0vst}, $T=1\;{\rm nK}$. The initial sample contains many phase singularities, including a few that are circulating counter to the cloud rotation. Counter-circulating vortices only tend to arise in the initial state, are thermodynamically very energetic, and are quickly annihilated by positive vortices during post-quench dynamics. The condensate, shown in Fig.~\ref{condBand_single_trajectory_sgpeRBEC}, forms with many vortices already present, but in a highly disordered configuration. The vortices then assemble into a regular triangular lattice.

At low temperatures the equilibrium state of the system consists of a an Abrikosov vortex lattice with small thermal fluctuations of the vortices about their average positions. The time dynamics for the highest temperature of Fig.~\ref{fig:N0vst}, $T=11\;{\rm nK}$, are shown in Fig.~\ref{second_single_trajectory_sgpeRBEC}. We see that growth proceeds into a disordered state for which periodic order remains frustrated by significant thermal fluctuations. 

\subsection{Equilibrium states}
\begin{figure}[tbp]
\centering
\includegraphics[width=7.5cm]{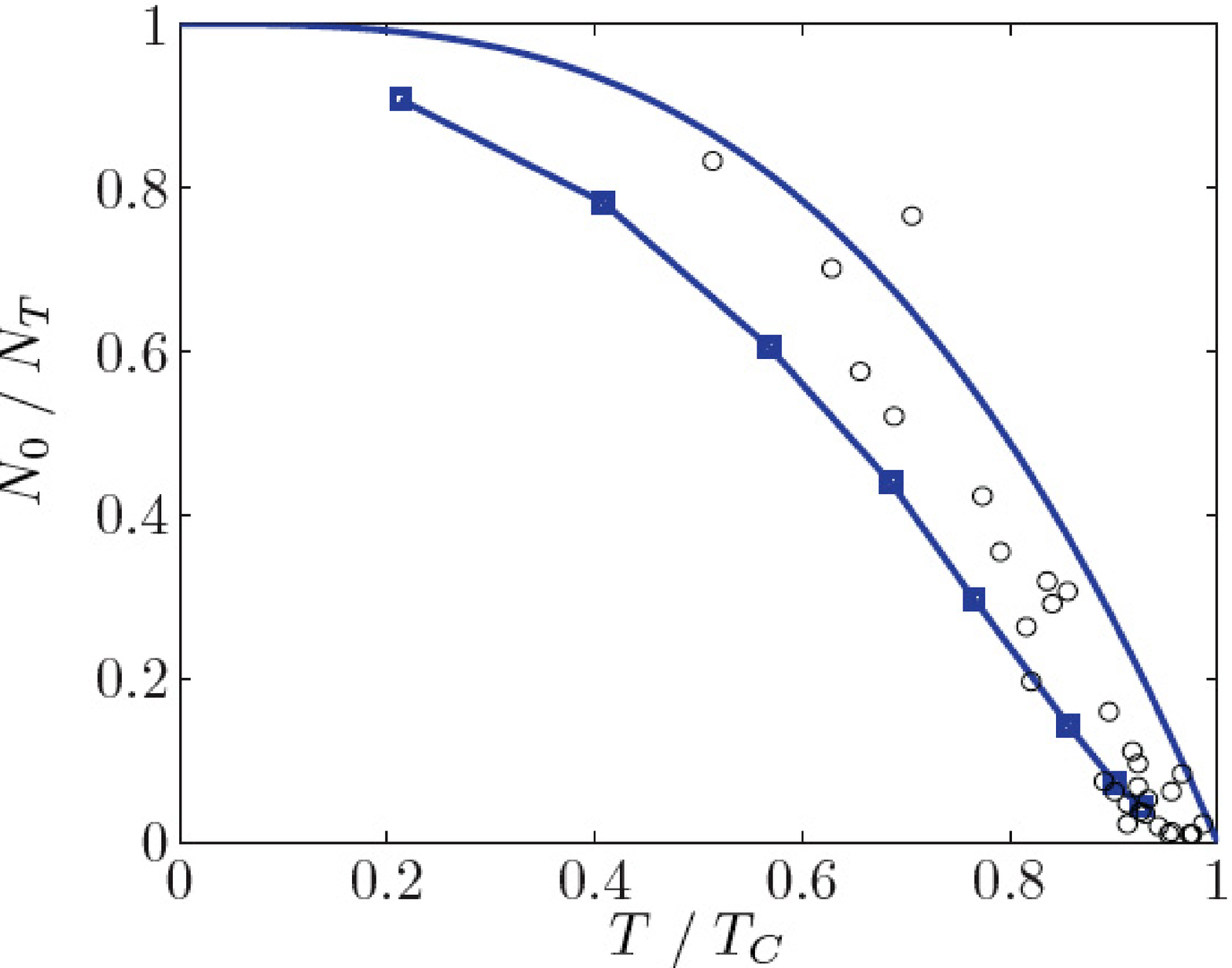}%
\caption{Equilibrium condensate fraction versus temperature. The boxes show the result of SGPE simulations for the rotating Bose gas at constant rotation frequency $\Omega=0.979\omega_r$. The solid line is the ideal gas expression, which is invariant with respect to $\Omega$. For comparison the open circles are the experimental data for the non-rotating system reported in Reference~\cite{Ensher1996a}.}
\label{fig:condFrac}
\end{figure}

We now consider equilibrium properties of the rotating system, in the grand canonical ensemble generated by evolution according to the SGPE. In Fig.~\ref{fig:condFrac} we plot the condensate fraction of the final state for the temperatures used in Fig.~\ref{fig:N0vst}. The total atom number in the system $N_T$ is calculated as $N_T= N_C+N_{NC}$, where $N_{NC}$ is given by Eq.~(\ref{NNC}) evaluated for our chosen energy cutoff, and $N_C$ is the mean total number of atoms in the condensate band. 

We note that the ideal gas relationship between condensate fraction and temperature, $N_0/N=1-(T/T_C)^3$, is unaltered by rotation, provided the rotating frame transition temperature, given by Eq.~\ref{Tc}, is used in place of the non-rotating transition temperature. The rotating ideal gas curve is plotted for comparison, and we see that the condensate fraction predicted by the SGPE simulations is suppressed relative to the ideal gas result. Although we are unable to simulate the nonrotating system with our two-dimensional algorithm, since the ideal gas behavior is universal with respect to rotation we also compare with the experimental data for the non-rotating Bose gas~\cite{Ensher1996a}. Our simulations show a condensate fraction significantly suppressed relative to the experimental data for the non-rotating system. 

This difference presumably comes about from two effects. The most obvious is the loss of long range order caused by thermal motion of vortices that are absent from the non-rotating system. A less obvious and possibly less significant effect is that the reduction of condensate number by interactions is more pronounced for a rotating gas, due to the exclusion of atoms from vortex cores. 
Our highest temperature simulation, at $T=11\;{\rm nK}$, is right on the edge of Bose-Einstein condensation, and this is reflected in the lack of long range order seen in Fig.~\ref{second_single_trajectory_sgpeRBEC}. 

\begin{figure}[tbp]
\centering
\includegraphics[width=7.5cm]{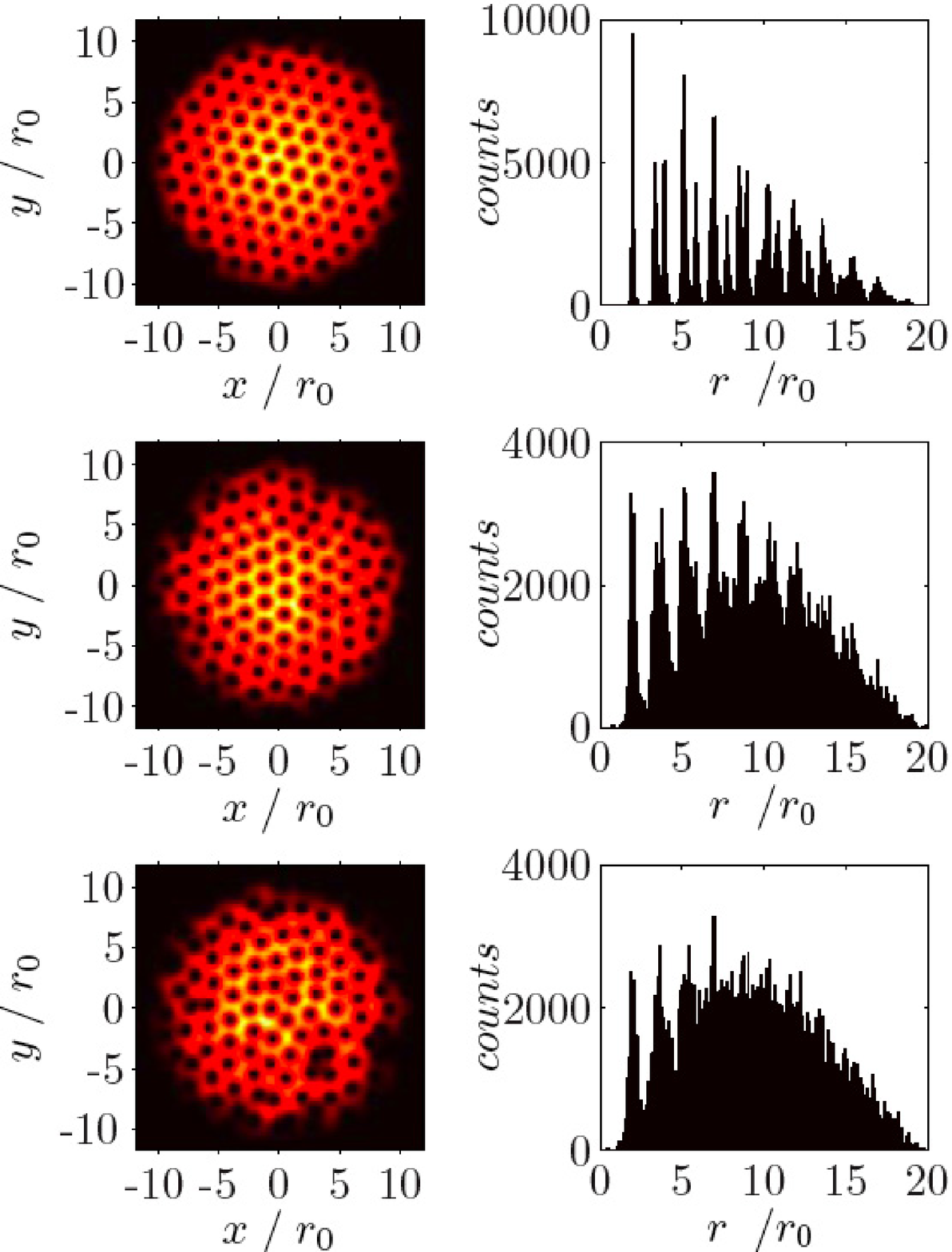}%
\caption{(color online) Condensate density and vortex position histograms. Top: $T=1\;{\rm nK}$, middle: $T=5\;{\rm nK}$, bottom: $T=11\;{\rm nK}$. The highly ordered Abrikosov lattice has its order degraded by thermal fluctuations for increasing temperature.}
\label{fig:vortHist}
\end{figure}

The loss of long range order can be seen more clearly by looking at the relative positions of vortices in the condensate wave-function at different temperatures~\cite{Engels2003a}. In Figure~\ref{fig:vortHist} we plot the condensate wave function, along with the histogram of the separation of each pair of vortices in the system. We only consider vortices with radius $r_v\leq 10 r_0$. For the lowest temperature ($T=1\;{\rm nK}$) the histogram is indistinguishable from the ground state histogram (not shown). The presence of many peaks for large separations reflects the long range spatial order in the system. As temperature increases, order is reduced starting at large vortex separation. At the highest temperature ($T=11\;{\rm nK}$) we see that only two histogram peaks clearly persist, corresponding to nearest and next-to nearest neighbor order being preserved.
\section{Conclusions}
We have developed and applied the SGPE theory to the rotating Bose gas.
The formalism presented here can be equally applied to the non-rotating case and in that limit extends the prior SGPE theory~\cite{SGPEI,SGPEII,SGPEIII} by providing expressions for the dissipation rates. We have developed an efficient algorithm for numerically integrating the SGPE in a rotating reference frame and applied it to the case of quenching a rotating Bose gas below threshold, modeling the rapid-quench limit of the experiment at JILA~\cite{Haljan2001}. 

Our simulations indicate that many vortices co-rotating with the thermal cloud are trapped in the emerging condensate, reflecting the large angular momentum of the system. In our calculations these primordial vortices are nucleated from vortices in the individual samples of the initial Wigner distribution and from fluctuations arising from the condensate growth terms in the SGPE. The process can be regarded as {\em stimulated} vortex production, in contrast with the non-rotating case where vortices can appear spontaneously but the total vorticity in the BEC must be zero on the average~\cite{Drummond1999}.
 
For low temperatures the vortices order into a regular Abrikosov lattice as the system approaches thermal equilibrium with the quenched thermal cloud. At sufficiently high temperatures lattice crystallization is frustrated by thermal fluctuations of the positions of vortex cores. This disorder is reflected in a loss of long range correlations in the relative positions of vortices that are most evident as the system approaches $T_C$. 

We note that the concept of rotation has not been investigated in the context of the Kibble-Zurek mechanism, which explains vortex nucleation during the phase transition in terms of spontaneous symmetry breaking and critical slowing down~\cite{Anglin1999}. 
Future quantitative studies of dynamics at the transition will allow for a deeper understanding of the condensation process, in particular the role of spontaneous symmetry breaking as a vortex nucleation mechanism in rotating Bose-Einstein condensation.

We gratefully acknowledge B. P. Anderson and M. K. Olsen for their critical reading of this manuscript, and P. Jain, P. B. Blakie and M. D. Lee for helpful discussions about this work. 

This research was supported by the Australian Research Council Center of Excellence for Quantum Atom Optics, by the New Zealand Foundation for Research Science and Technology under the contract NERF-UOOX0703: Quantum Technologies, and by the Marsden Fund under contracts PVT202 and UOO509.
\appendix
\section{Calculation of dissipation rates}
\label{app:rates}
Reservoir interaction rates closely connected to those required here have been previously computed within the context of the quantum kinetic theory (QKT) of BEC~\cite{QKV}. We now derive the form of the growth and scattering rates including the effects of the cutoff, the external trapping potential and the rotation of the thermal cloud. 
\subsection{Growth}
\subsubsection{General form}
From Eq.~(84) of Ref.~\cite{SGPEII}, the growth amplitude is
\begin{equation}
G(\mbf{x})\equiv\int d^3\mbf{v}\;G^{(\scriptscriptstyle +)}(\mbf{x},\mbf{v},0)=G_1(\mbf{x})+G_2(\mbf{x})
\label{GdefApp}
\end{equation}
where
\begin{equation}
 \begin{split}
G_1(\mbf{x})\equiv&\frac{u^2}{(2\pi)^5\hbar^2}\int\int\int_{R_{NC}}d^3\mbf{K}_1d^3\mbf{K}_2d^3\mbf{K}_3\\
&\times F(\mbf{x},\mbf{K}_1)F(\mbf{x},\mbf{K}_2)\Delta_{123}(0,0),
\end{split}
\end{equation}
\begin{equation}
\begin{split}
G_2(\mbf{x})\equiv&\frac{u^2}{(2\pi)^5\hbar^2}\int\int\int_{R_{NC}}d^3\mbf{K}_1d^3\mbf{K}_2d^3\mbf{K}_3\\
&\times F(\mbf{x},\mbf{K}_1)F(\mbf{x},\mbf{K}_2)F(\mbf{x},\mbf{K}_3)\Delta_{123}(0,0),
\end{split}
\end{equation}
and $\Delta_{123}(\mbf{k},\epsilon)\equiv\delta(\mbf{K}_1+\mbf{K}_2-\mbf{K}_3-\mbf{k})\delta(\omega_1+\omega_2-\omega_3-\epsilon/\hbar)$ conserves energy and momentum. Note that effect of working in the rotating frame is only to introduce the effective potential term $-m\Omega^2 r^2/2$ into the cut-off and Wigner functions in $G({\bf x})$, which is otherwise unaltered.
We will take the reservoir to be in equilibrium with Bose-Einstein distribution at chemical potential $\mu$ and temperature $T$. Integration over the momentum $\delta$-function reduces $G_1(\mbf{x})$ to
\begin{equation}
\begin{split}
G_1(\mbf{x})=&\frac{u^2}{(2\pi)^5\hbar^2}\frac{8m\pi^2}{\hbar}\sum_{p,q=1}^\infty e^{\beta(\mu-V_{\rm eff}(\mbf{x}))(p+q)}\int_{K_R(\mbf{x})}^\infty K_1dK_1\\
&\times\int_{K_R(\mbf{x})}^\infty K_2dK_2\;e^{-(pK_1^2+qK_2^2)\beta\hbar^2/2m}\\
&\times \int_{-1}^1 dz\;\delta(z-mV_{\rm eff}(\mbf{x})/\hbar^2K_1K_2),
\end{split}
\end{equation}
where Eq.~(\ref{rspec}) gives the expression for the momentum cutoff 
\begin{equation}
K_R(\mbf{x})=\sqrt{2m[E_R-V_{\rm eff}(\mbf{x})]/\hbar^2}
\end{equation}
Changing to variables $s=p\beta\hbar^2K_1^2/2m$, $t=q\beta\hbar^2K_2^2/2m$, and taking account of the delta-function, we obtain
\begin{equation}
\begin{split}
G_1(\mbf{x})=&\frac{u^2}{(2\pi)^5\hbar^2}\frac{8m\pi^2}{\hbar}\left(\frac{m}{\beta\hbar^2}\right)^2\sum_{p,q=1}^\infty \frac{e^{\beta(\mu-V_{\rm eff}(\mbf{x}))(p+q)}}{pq}\\
&\times\int_{s_{min}(\mbf{x})}^\infty ds\;e^{-s}\int_{t_{min}(\mbf{x},s)}^\infty  dt\;e^{-t},
\end{split}
\end{equation}
where $s_{min}(\mbf{x})=p\beta[E_R-V_{\rm eff}(\mbf{x})]$, and $t_{min}(\mbf{x},s)={\rm min}\left\{q\beta [E_R-V_{\rm eff}(\mbf{x})], (\beta V_{\rm eff}(\mbf{x}))^2pq/4s\right\}$. 
We now show that the rate is composed of two regions: an inner region where the rate is spatially invariant, and an outer limit where there is a weak spatial dependence. 
\subsubsection{Inner region}
From the from of the lower limits we find that wherever $V_{\rm eff}(\mbf{x})\leq 2E_R/3$,  $t_{min}(\mbf{x},s)=s_{min}(\mbf{x})=q\beta [E_R-V_{\rm eff}(\mbf{x})]$. Proceeding similarly for $G_2(\mbf{x})$ we find the same condition to hold, so that in the inner region the growth rates are spatially invariant, taking the final form
\begin{eqnarray}
G_1^{in}&=&\frac{4m}{\pi\hbar^3}(ak_BT)^2\left[\ln{\left(1-e^{\beta(\mu-E_R)}\right)}\right]^2,\\\label{G1in}
G_2^{in}&=&\frac{4m}{\pi\hbar^3}(ak_BT)^2e^{2\beta(\mu-E_R)}\nonumber\\
&&\times\sum_{r=1}^\infty\;e^{r\beta(\mu-2E_R)}\left(\Phi[e^{\beta(\mu-E_R)},1,r+1]\right)^2,\label{G2in}
\end{eqnarray}
where the {\em Lerch transcendent} is defined as $\Phi[z,s,a]=\sum_{k=0}^\infty z^k/(a+k)^s$. 
These terms are superficially similar to the condensate growth rates previously obtained using the quantum kinetic theory~\cite{QKPRLII}, although the form of the cutoff dependence is different in the SGPE theory. Note that this form is independent of any collisional contribution to the effective potential.

\subsubsection{Outer region}
In the region where $ 2E_R/3\leq V_{\rm eff}(\mbf{x})\leq E_R$ the rates depend on position, albeit quite weakly. At the edge of the condensate band where $V_{\rm eff}(\mbf{x})\equiv E_R$, we find
\begin{eqnarray}
G_1^{out}&=&\frac{4m}{\pi\hbar^3}(ak_BT)^2\beta E_R\nonumber\\
&&\times\sum_{p,q=1}^\infty\frac{e^{\beta(\mu-E_R)(p+q)}}{\sqrt{pq}}K_1(\beta\sqrt{pq})\\
G_2^{out}&=&\frac{4m}{\pi\hbar^3}(ak_BT)^2\beta E_R,\nonumber\\
&&\times\sum_{p,q,r=1}^\infty\frac{e^{\beta(\mu-E_R)(p+q)+\beta(\mu-2E_R)r}}{\sqrt{(p+r)(q+r)}}\nonumber\\
&&\times K_1(\beta\sqrt{(p+r)(q+r)}),
\end{eqnarray}
where $K_\alpha(x)$ is the modified Bessel function of the second kind.

\begin{figure}[tbp]
\centering
\includegraphics[width=8.5cm]{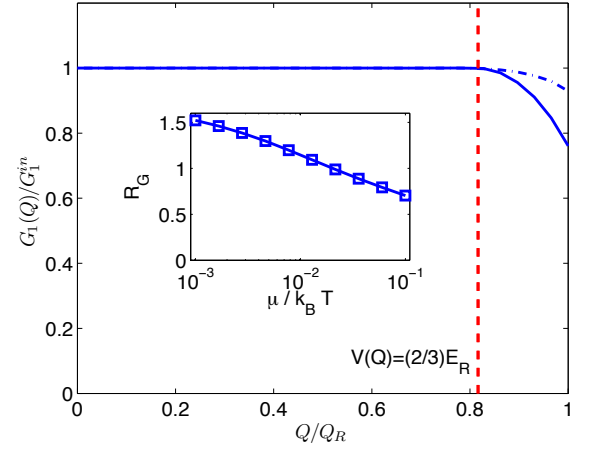}
\caption{Spatial variation of $G_1(Q)$ in a spherical trap $V(Q)=m\omega_r^2 Q^2/2$. Growth rates are shown for $\mu=0.1 k_BT$ (solid line) and $\mu = 0.01 k_BT$ (chain line). The dashed vertical line indicates the point $V(Q)=2E_R/3$ where the rates begin to vary. The condensate band terminates at $Q_R=\sqrt{2E_R/m\omega_r^2}$. The ratio $R_G$ (see text) is shown versus $\mu/k_B T$ (inset).}
\label{gvar}
\end{figure}

In Fig.~\ref{gvar} we compute the variation of $G_1(\mbf{x})$ with radius for a spherical parabolic potential. For illustration we have also neglected the collisional contribution to the effective potential. To show the maximum variation of the total rate we also plot (inset) the ratio $R_G=(G_1^{out}+G_2^{out})/(G_1^{in}+G_2^{in})$ versus $\mu/k_B T$. For all data we use $E_R=3\mu$ which is typically a good approximation for the energy at which the spectrum of the many body system becomes single particle-like~\cite{SGPEII}. The rates are spatially invariant except for a small region near the edge of the condensate band where the scattering phase space in the harmonic trap is modified by the energy cutoff. Since the variation with radius is usually not significant, for our simulations we will use the approximate growth rate 
\begin{equation}\label{approxG}
G(\mbf{x})\approx\gamma\equiv G_1^{in}+G_2^{in},
\end{equation}
which is exact in the inner region and a good approximation elsewhere. 

\subsection{Scattering}
Our starting point of the scattering amplitude calculation, that is related to the actual scattering via Fourier transform, is Eq.~(86) of Ref.~\cite{SGPEII} 
\begin{equation}\label{MtildeDef}
\begin{split}
\tilde{M}(\mbf{x},\mbf{k},\epsilon)=&\frac{4\pi u^2}{\hbar^2}\int_{R_{NC}} \frac{d^3\mbf{K}_1}{(2\pi)^3}\int_{R_{NC}} \frac{d^3\mbf{K}_2}{(2\pi)^3}\;\Delta_{12}(\mbf{k},\epsilon)\\
&\times F(\mbf{x},\mbf{K}_1)[1+F(\mbf{x},\mbf{K}_2)],
\end{split}
\end{equation}
where $\Delta_{12}(\epsilon,\mbf{k})\equiv\delta(\omega_1+\omega_2-\epsilon/\hbar)\delta(\mbf{K}_1+\mbf{K}_2-\mbf{k})$.
Upon evaluating the momentum delta-function we find
\EQ{
&&\tilde{M}(\mbf{x},\mbf{k},0)=\frac{4mu^2}{(2\pi)^5\hbar^3}\int_{R_{NC}}d^3\mbf{K}\;F(\mbf{x},\mbf{K})[1+F(\mbf{x},\mbf{K})]\nonumber\\
&&\;\;\;\;\;\hspace{2cm}\times\delta\left(2\left\{\mbf{K}-\frac{m}{\hbar}\mbf{\Omega}\times\mbf{x}\right\}\cdot\mbf{k}-\mbf{k}^2\right).
}
Using the notation $\hbar\omega(\mbf{x},\mbf{K})=e_{\mbf{K}-m\mbf{\Omega}\times\mbf{x}/\hbar}(\mbf{x})$, where
\EQ{\label{edef}
e_{\mbf{K}}(\mbf{x})\equiv\frac{\hbar^2\mbf{K}^2}{2m}+V_{\rm eff}(\mbf{x}),
} 
in the variable $\mbf{q}=\mbf{K}-m(\mbf{\Omega}\times\mbf{x})/\hbar$ we have
\EQ{\label{Mterm}
\tilde{M}(\mbf{x},\mbf{k},0)&=&\frac{4mu^2}{(2\pi)^5\hbar^3}\int_{R_{NC}}d^3\mbf{q}\;\delta\left(2\mbf{q}\cdot\mbf{k}-\mbf{k}^2\right),\nonumber\\
&&\times f_{T,\mu}(e_{\mbf{q}}(\mbf{x}))[1+f_{T,\mu}(e_{\mbf{q}}(\mbf{x}))],
}
where $f_{T,\mu}(z)=[e^{(z-\mu)/k_BT}-1]^{-1}$. Writing $\tilde{M}(\mbf{x},\mbf{k},0)=\tilde{M}_1(\mbf{x},\mbf{k})+\tilde{M}_2(\mbf{x},\mbf{k})$ to distinguish the term with one or two BE-distributions as before, and choosing the $z$-component of $\mbf{K}_1$ along $\mbf{k}$, we find, for the term involving one BE-distribution
\begin{equation}
\begin{split}
\tilde{M}_1(\mbf{x},\mbf{k})=&\frac{4mu^2}{(2\pi)^5\hbar^3}\frac{\pi}{|\mbf{k}|}\sum_{p=1}^\infty e^{p\beta(\mu-V_{\rm eff}(\mbf{x}))}\\
&\times \int_{K_R(\mbf{x})}^\infty K_1 dK_1 e^{-p\beta \hbar^2 K_1^2/2m}\Theta (|\mbf{k}|/2K_1\leq 1).
\end{split}
\end{equation}
The theta function generates the effective cutoff $K_1\geq K_{min}(\mbf{x})\equiv {\rm max}\{K_R(\mbf{x}),|\mbf{k}|/2\}$. Evaluating the integral gives
\begin{equation}
\begin{split}
\tilde{M}_1(\mbf{x},\mbf{k})=&\frac{4mu^2}{(2\pi)^5\hbar^2}\frac{\pi m}{\beta \hbar^2|\mbf{k}|}\\\label{MoneTilde}
&\times \frac{e^{\beta(\hbar^2K_{min}^2(\mbf{x})/2m+V_{\rm eff}(\mbf{x})-\mu)}}{\left(e^{\beta(\hbar^2K_{min}^2(\mbf{x})/2m+V_{\rm eff}(\mbf{x})-\mu)}-1\right)^2},
\end{split}
\end{equation}
which can be further simplified as follows.
We want the condition for $K_{min}(\mbf{x})=K_R(\mbf{x})$, which can be expressed as $\hbar^2\mbf{k}^2/8m\leq E_R-V_{\rm eff}(\mbf{x})$. The restriction of $|\mbf{k}|$ is determined by the requirement that $\mbf{k}=\mbf{k}_1+\mbf{k}_2$, where $\hbar^2\mbf{k}_{1,2}^2/2m+V_{\rm eff}(\mbf{x})\leq E_R$ for atoms in the condensate band. Then $\hbar^2\mbf{k}^2/8m= \hbar^2 \mbf{k}_1^2/8m+\hbar^2\mbf{k}_2^2/8m+\hbar^2 \mbf{k}_1\cdot\mbf{k}_2/4m\leq E_R-V_{\rm eff}(\mbf{x})$ and the condition always holds. The exponential argument in Eq. (\ref{MoneTilde}) then simplifies to $\beta(E_R-\mu)$. The term involving two BE-distributions, $\tilde{M}_2(\mbf{x},\mbf{k})$, is computed similarly, and finally we arrive at 
\begin{equation}
\tilde{M}(\mbf{x},\mbf{k},0)=\frac{16\pi a^2 k_B T}{(2\pi)^3\hbar |\mbf{k}|}\frac{e^{\beta(E_R-\mu)}}{\left(e^{\beta(E_R-\mu)}-1\right)^2}\equiv \frac{\cal M}{(2\pi)^3|\mbf{k}|}.
\label{Mfinal}
\end{equation}
As the result is independent of $\mbf{x}$, in this paper we drop the redundant arguments and in place of $M(\mbf{u},\mbf{v},0)$, we obtain the the scattering amplitude given by the Fourier transform of Eq.~(\ref{Mfinal})
\begin{equation}
M(\mbf{v})=\frac{{\cal M}}{(2\pi)^{3}}\int d^3\mbf{k}\frac{e^{-i\mbf{k}\cdot\mbf{v}}}{ |\mbf{k}|}.
\end{equation}
\section{Numerical algorithm for the PGPE in a rotating frame}
\label{app:numerics}
In this appendix we outline our numerical method for evolving the two-dimensional SGPE in cylindrical coordinates in a rotating frame of reference. For our purposes the method has several advantages over Fourier or Crank-Nicholson methods. It is a pseudo-spectral method  makes use of a mixture of Gauss-Laguerre and Fourier quadratures for efficiency, and most importantly, allows an energy cutoff to be efficiently implemented in the rotating frame. 

Since the noise in the simple growth SGPE is additive, we use the stochastic Runge-Kutta algorithm due to the superior stability properties of Runga-Kutta methods ~\cite{Milstein}. Consequently, our central requirement is a means to efficiently evaluate $\PP L_{\rm GP}\alpha(\mbf{x},t)$, or the projected Gross-Pitaevskii equation
\EQ{\label{RFgpe}i\frac{\partial\alpha}{\partial
t}=\PP\left\{\left[-\frac{1}{2}\nabla_\perp^2+i\Omega\partial_\theta+\frac{r^2}{2} + \lambda|\alpha|^2\right]\alpha\right\},
} 
where $\nabla_\perp^2$ is the two dimensional Laplacian. The deterministic part of the SGPE evolution is then found by transforming to an appropriate rotating frame. The solutions of the linear Schr\"{o}dinger equation
\EQ{\label{RFeqn} \left(-\frac{1}{2}\nabla_\perp^2+i\Omega 
\partial_\theta+\frac{r^2}{2}\right)Y_{nl}(r,\theta)=\omega_{nl}^RY_{nl}(r,\theta),}
are the Laguerre-Gaussian modes
\EQ{\label{RFeigs}Y_{nl}(r,\theta)=\sqrt{\frac{n!}{\pi(n+|l|)!}}e^{il\theta}r^{|l|}e^{-r^2/2}L_n^{|l|}(r^2),
}
where $L_n^m(x)$ are the associated Laguerre polynomials, and $n$ and $l$ are the quantum numbers for radial and angular excitations. The spectrum of the modes is
\EQ{\label{RFspec}\omega_{nl}^R=2n+|l|-\Omega l +1,}
where $n\in\{0, 1, 2, \dots\}\;l\in\{\dots,-1, 0, 1, 2, \dots\}$.
\par
Imposing an energy cut-off in the rotating frame causes the range of available energy states to be biased toward states rotating in the same direction as the frame.
We define
the energy cut-off in the rotating frame by keeping all modes satisfying
\EQ{\label{lagspec}
2n+|l|-\Omega l\leq \bar{N},}
corresponding to cut-off energy $E_R=\hbar\omega_r(\bar{N}+1)$.
From Eq. (\ref{lagspec}), the quantum numbers are restricted by the cutoff to
\EQ{\label{nrange}
0\leq &n&\leq\left[\bar{N}/2\right]\\
\label{mrange}
-l_{\scriptscriptstyle-}(n)\leq
&l& \leq l_{\scriptscriptstyle+}(n),
} 
where in this appendix $[x]$ represents the integer part of $x$, and $l_\pm(n)=\left[(\bar{N}-2n)(1\mp\Omega)\right]$. The projected wave-function then takes the form
\EQ{\label{Lpsi}
\alpha(r,\theta)=\sum_{n=0}^{[\bar{N}/2]}\sum_{l=-l_{\scriptscriptstyle-}(n)}^{l_+(n)}\alpha_{nl}(t)Y_{nl}(r,\theta).
}
Projecting (\ref{RFgpe}) onto the single particle basis (\ref{RFeigs}) leads to the
equation of motion
\EQ{ i\frac{\partial
\alpha_{nl}}{\partial \tau}=\omega_{nl}^R\alpha_{nl}+\lambda F_{nl}(\alpha),
}
where 
\EQ{
F_{nl}(\alpha)=\int_0^{2\pi}d\theta\int_0^\infty rdr\;Y_{nl}^*(r,\theta)|\alpha(r,\theta)|^2\alpha(r,\theta)
}
is the projection of the GPE nonlinear term, which is the numerically difficult term to evaluate for any pseudo-spectral method of the solving the GPE.
Changing variables gives
\EQ{\label{Frf}F_{nl}(\alpha)&=&\frac{1}{2\pi}\int_0^{2\pi}d\theta\;e^{-il\theta}\int_0^\infty dx\;\Phi_{nl}(x)\pi\nonumber\\
&&\times|\alpha(\sqrt{x},\theta)|^2\alpha(\sqrt{x},\theta)\;\;}
where $\Phi_{nl}(x)=Y_{nl}(\sqrt{x},\theta)e^{-il\theta}$. 
The integral to evaluate at every time step of propagation is always of the form
\EQ{\label{Idef}
I=\frac{1}{2\pi}\int_0^{2\pi}d\theta\;e^{-il\theta}\int_0^\infty dx\;e^{-2x}Q(x,\theta)}
where $Q(x,\theta)$ is a polynomial in $x$ and $e^{i\theta}$ of order determined by the cutoff.
The angular integration can therefore be carried out efficiently using an FFT. The order of $e^{i\theta}$ in $\alpha(\mbf{x})$ is
constrained by (\ref{mrange}) to be $-\bar{l}_-\equiv-l_{\scriptscriptstyle-}(0)\leq
l \leq l_{\scriptscriptstyle+}(0)\equiv\bar{l}_+$.
Choosing the $\theta$ grid so as to accurately compute all integrals $\frac{1}{2\pi}\int_0^{2\pi}d\theta\; e^{-ilq}$, 
where $q$ is an integer in the range $-2(\bar{l}_{\scriptscriptstyle-}+\bar{l}_{\scriptscriptstyle+})\leq q\leq 2(\bar{l}_{\scriptscriptstyle-}+\bar{l}_{\scriptscriptstyle+})$, the integral can be carried out exactly for the condensate band by discretising $\theta$ as $\theta_j=j\Delta\theta=j2\pi/N_\theta$ for $j=0,1,\dots
,N_\theta-1$, where $N_\theta \equiv2(\bar{l}_{\scriptscriptstyle-}+\bar{l}_{\scriptscriptstyle+})+1$,
to give
\EQ{\int_0^{2\pi}d\theta\;e^{-il\theta}Q(x,\theta)
=\frac{2\pi}{N_\theta}\sum_{j=0}^{N_\theta-1}e^{-i2\pi
jl/N_\theta}Q(x,\theta_j).}
The $x$ integral is computed by noting that the polynomial part of the wave-function is of maximum order 
$
\left[\bar{N}/(1-\Omega)\right]/2
$
in $x$ (since $x^{|l|/2}L_n^{|l|}(x)$ is a polynomial of degree $n+|l|/2\leq |\bar{l}_{+}|/2$) and so the
integrand is of maximum order 
$
2\left[\bar{N}/(1-\Omega)\right],
$
which can be computed exactly using a
Gauss-Laguerre quadrature rule of order 
$N_x\equiv\left[\bar{N}/(1-\Omega)\right]+1$.
The radial integral in (\ref{Idef}) is 
\EQ{\int_0^\infty\;dx\;Q(x,\theta)e^{-2x}=\frac{1}{2}\sum_{k=1}^{N_x}\;w_k\;Q(x_k/2,\theta),}
where $x_k$ are the roots of $L_{N_x}(x)$ and the weights $w_k$ are known constants \cite{Cohen1973}.
We can thus write (\ref{Frf}) as
\begin{equation}
\begin{split}
F_{nl}(\alpha)=&\sum_{k=1}^{N_x}\tilde{w}_k\;\Phi_{nl}(x_k/2)\;\frac{1}{N_\theta}\sum_{j=0}^{N_\theta-1}e^{-i2\pi
jl/N_\theta}\\
&\times|\alpha(\sqrt{x_k/2},\theta_j)|^2\alpha(\sqrt{x_k/2},\theta_j),
\end{split}
\end{equation}
where the combined weight variable is
$\tilde{w}_k=\pi w_k e^{x_k}/2$.
Once the modes are pre-computed at the Gauss points
\EQ{P_{kn}^{|l|}=\Phi_{n|l|}(x_k/2)}
the mixed quadrature spectral Galerkin method for evolving the GPE is
\begin{itemize}
\item[(i)] Compute the radial transform
\EQ{
\chi_{kl}=\sum_{n=0}^{[\bar{N}/2]}P_{kn}^{|l|}\;\alpha_{nl}.
}
\item[(ii)] Construct the position field at the quadrature points
$\Psi_{kj}$ using the FFT, after zero padding $\chi_{kl}$ to length $N_\theta$ in the $l$ index, $\tilde{\chi}_{kl}\equiv{\rm pad}_l(\chi_{kl},N_\theta)$:
\EQ{\Psi_{kj}=\alpha(\sqrt{x_k/2},\theta_j)=\sum_{l=0}^{N_\theta-1}\tilde{\chi}_{k,l}\;e^{i2\pi jl/N_\theta}.}
\item[(iii)] Compute the nonlinear term
\EQ{\Xi_{kj}=|\Psi_{kj}|^2\Psi_{kj}.}
\item[(iv)] Compute the IFFT
\EQ{\Theta_{kl}=\frac{1}{N_\theta}\sum_{j=0}^{N_\theta-1}\Xi_{kj}\;e^{-i2\pi
jl/N_\theta}.
}
\item[(v)] Calculate the Gauss-Laguerre quadrature
\begin{equation}
F_{nl}(\alpha)=\sum_{k=1}^{N_x}\;\tilde{w}_k\;\Theta_{kl}\;P_{kn}^{|l|}.
\end{equation}
\end{itemize}

We emphasize that the transform matrix $P_{kn}^{|l|}$ is separable, but requires a different $kn$ matrix for each $|l|$. The extra memory required to store $P_{kn}^{|l|}$ (cf. the method of \cite{Dion2003} for the nonrotating GPE) is mitigated by the modest memory used by the FFT. The order of the FFT rule required is consistent with the one dimensional harmonic oscillator rule given in \cite{Dion2003} where it was shown that to numerically evolve a Gaussian weighted polynomial wave-function of order $\bar{N}$ in $x$ according to the PGPE, a Gaussian quadrature rule of order $2\bar{N}+1$ is required. Since the FFT is a quadrature rule for unweighted polynomials in $e^{i\theta}$ we can expect the same behavior. This is evident once we note that $\bar{l}_-+\bar{l}_+$ is the order of the wave-function in powers of $e^{i\theta}$ and therefore plays the same role as $\bar{N}$ in the standard Gauss-Hermite quadrature. 

The discretisation of angle and radius have thus been chosen as the smallest number of points needed to faithfully project the nonlinear term onto the representation basis without numerical error. The computational load of the method scales as $O(N N_x N_\theta)$ where $N={\rm max}(N_x, \log{N_\theta})$, affording a significant advantage over Cartesian grid methods in the rapidly rotating regime since Cartesian grids are comparatively inefficient at representing the quantum states involved. Most notably, in the limit $\Omega\to\omega_r$ we may truncate the basis at the lowest Landau level $N_x= 1$ and the method reaches an optimal efficiency of one dimensional FFT.
\bibliographystyle{prsty}

\end{document}